\documentclass[]{aastex701}
\usepackage{amsmath}

\begin{document}

\title{Mode Composition Shapes Magnetic Anisotropy in Solar Wind Turbulence}

\author[orcid=0000-0003-4268-7763]{Siqi Zhao}
\affiliation{Deutsches Elektronen-Synchrotron DESY, Platanenallee 6, 15738, Zeuthen, Germany}
\affiliation{Institut für Physik und Astronomie, Universität Potsdam, D-14476, Potsdam, Germany}
\email{siqi.zhao@desy.de}  

\author[orcid=0000-0003-2560-8066]{Huirong Yan} 
\affiliation{Deutsches Elektronen-Synchrotron DESY, Platanenallee 6, 15738, Zeuthen, Germany}
\affiliation{Institut für Physik und Astronomie, Universität Potsdam, D-14476, Potsdam, Germany}
\email[show]{huirong.yan@desy.de}  

\author[orcid=0000-0003-1778-4289]{Terry Z. Liu}
\affiliation{Department of Earth, Planetary, and Space Sciences, University of California, LA, USA}
\email[show]{terryliuzixu@ucla.edu}  

\author[orcid=0000-0001-7205-2449]{Chuanpeng Hou}
\affiliation{Institut für Physik und Astronomie, Universität Potsdam, D-14476, Potsdam, Germany}
\email{terryliuzixu@ucla.edu}  

\correspondingauthor{Huirong Yan; Terry Z. Liu}

\begin{abstract}
Turbulence is a ubiquitous process that transfers energy across many spatial and temporal scales, thereby influencing particle transport and heating. Recent progress has improved our understanding of the anisotropy of turbulence with respect to the mean magnetic field; however, its exact form and implications for magnetic topology and energy transfer remain unclear. In this study, we investigate the nature of magnetic anisotropy in compressible magnetohydrodynamic (MHD) turbulence within low-$\beta$ solar wind using measurements from the Cluster spacecraft. By decomposing small-amplitude fluctuations into Alfvén and compressible modes, we reveal that magnetic anisotropy is largely mode dependent: Alfvénic fluctuations are broadly distributed in propagation angle, whereas compressible fluctuations are concentrated near the quasi-parallel (slab) direction, a feature closely linked to collisionless damping of compressible modes. For $\beta\rightarrow0$, compressible modes become dominant within the slab component at smaller scales. These findings advance our understanding of magnetic anisotropy in solar wind turbulence and offer a new perspective on the three-dimensional turbulence cascade, with broad implications for particle transport, acceleration, and magnetic reconnection.



\end{abstract}


\section{Introduction}





Plasma turbulence regulates the transfer of energy across a broad range of scales and plays a crucial role in astrophysical and space phenomena, such as star formation, solar and stellar coronal heating, solar wind heating and acceleration, and the transport of energetic particles \citep{McKee2007,Yan2002, Suzuki2006,Yan20082,Bruno2013, Yan2022, Zhao2025_si}. For over three decades, the anisotropy of turbulence with respect to the mean magnetic field ($\mathbf{B}_0$) has been widely recognized as a key feature of plasma turbulence. However, its detailed structure and its impact on magnetic topology and energy transfer remain unclear. Here, we report observational evidence directly linking magnetic anisotropy to mode composition in compressible magnetohydrodynamic (MHD) turbulence.

Extensive simulations and observations have demonstrated that the large-scale behavior of plasma turbulence can be described using incompressible or nearly incompressible MHD frameworks \citep{Montgomery1981, Matthaeus1990, Zank1992, Zank1993, Zank2017, Goldreich1995, Zhao2022, Zhao2025}. These models consistently reveal pronounced magnetic anisotropy in variance, power, wavevector distribution, spectral index, and energy transfer rate \citep{Horbury2012,Oughton2015}. A widely used representation is the two-component `slab+2D' model, in which Alfvén waves correspond to slab modes propagating along $\mathbf{B}_0$, whereas the two-dimensional (2D) component consists of fluctuations with wavevectors ($\mathbf{k}$) quasi-perpendicular to $\mathbf{B}_0$ \citep{Matthaeus1990,Zank1992,Zank1993,Zank2017}. A more comprehensive description is provided by the critical balance model, which characterizes strong turbulence across the full three-dimensional (3D) $\mathbf{k}$-space spectrum \citep{Goldreich1995}. The model predicts that the energy cascade follows scaling $k_\parallel \propto k_\perp^{2/3}$, so that turbulence becomes increasingly anisotropic at smaller scales, with correlation lengths much longer parallel to $\mathbf{B}_0$ than perpendicular to it \citep{Goldreich1995}. Here $k_\parallel$ and $k_\perp$ are wavenumbers parallel and perpendicular to $\mathbf{B}_0$, respectively.

Notably, most astrophysical and space plasmas with finite $\beta$, defined as the ratio of proton thermal to magnetic pressure, are inherently compressible, and compressibility plays a crucial role in their dynamics \citep{Hnat2005,Makwana2020,Zhang2020,Zhao2024a}. Understanding anisotropy in compressible turbulence is complicated by two key challenges. First, the energy cascade depends on mode composition, with each mode exhibiting distinct cascade behaviors. Alfvén and slow (magnetosonic) modes follow the critical-balance scaling $k_\parallel\propto k_\perp^{2/3}$, whereas fast (magnetosonic) modes show isotropy with a scaling resembling acoustic turbulence \citep{Zhao2024a,Zhao2024b,Hou2025}. Second, compressible modes undergo strong damping even at MHD scales \citep{YLD2004,Suzuki2006,Hou2025}. Damping quenches quasi-parallel slow modes, and it suppresses fast modes with high obliquity \citep{Zhao2024a}. Consequently, mode composition and damping jointly shape the anisotropy of compressible turbulence, a key to realistic plasma turbulence modeling.

Over the past three decades, significant progress has been made in decomposing small-amplitude fluctuations, for which nonlinear terms are much smaller than linear terms (i.e., $\delta B^2\ll\delta \mathbf{B}\cdot \mathbf{B}_0$), where $\delta B$ denotes the amplitude of magnetic fluctuations. In a homogeneous plasma with a uniform $\mathbf{B}_0$, small-amplitude fluctuations can be represented as a superposition of three linear MHD eigenmodes: incompressible Alfvén modes, and compressible fast and slow modes \citep{Glassmeier1995,Cho2002,Zhao2022}. More recently, \cite{Zank2023} developed a new linear mode decomposition method that highlights the role of nonpropagating modes, including the advected entropy mode and the advected magnetic-island (flux-rope) mode. Applied to solar-wind turbulence \citep{Zank2024} and interplanetary shocks \citep{Gautam2025}, this framework demonstrates that entropy modes contribute notably to density fluctuations, whereas magnetic-island structures make a non-negligible contribution to transverse magnetic fluctuations.

In this study, we investigate magnetic anisotropy
by decomposing compressible magnetic fluctuations in low-$\beta$ solar wind. In the low-$\beta$ regime, compressible magnetic energy is dominated by fast modes \citep{Zhao2021}, whose damping is well described analytically \citep{Yan2004,Petrosian2006} and confirmed by kinetic simulations \citep{Hou2025}, allowing a quantitative evaluation of the underlying process. The paper is organized as follows: Section 2 describes the data sets and methodology; Section 3 presents the observations; and Section 4 discusses and summarizes the key findings.

\section{Data and Methodology}

We used data from Cluster-1 for the angular analysis and from all four spacecraft for the wavenumber analysis. The magnetic field data were obtained from the Fluxgate Magnetometer (FGM) \citep{Balogh1997} at $22.5$ Hz, and proton bulk velocity from Cluster Ion Spectrometry’s Hot Ion Analyzer (CIS-HIA) \citep{Reme2001} at $4$ s cadence. Additional proton parameters, notably the more accurate proton temperature, were taken from the OMNI dataset at a $1$-min resolution, as CIS-HIA ion moments are less reliable in the solar wind.

\begin{figure*}[ht!]
\centering
\includegraphics[scale=0.5]{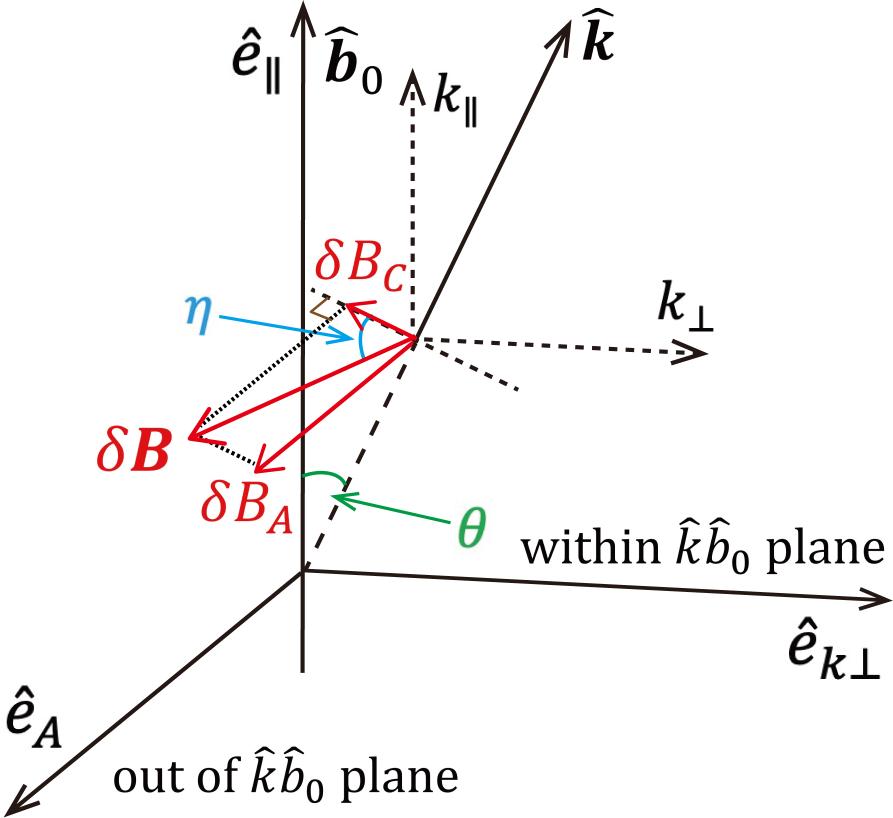}
\caption{\label{fig:1} Magnetic fluctuations in the $\hat{k}\hat{b}_0$ coordinates. $\theta$ is the angle between $\hat{\mathbf{k}}$ and $\hat{\mathbf{b}}_0$, and $\eta$ is the angle between $\mathbf{\delta B}$ and the $\hat{k}\hat{b}_0$ plane, estimated as $\eta=arctan(\sqrt{P_{A}/P_{C}})$.}
\end{figure*}

As illustrated in Fig. \ref{fig:1}, small-amplitude magnetic fluctuations were analyzed in the $\hat{k}\hat{b}_0$ coordinate defined by the unit $\hat{\mathbf{k}}$ and $\hat{\mathbf{b}}_0=\mathbf{B}_0/|\mathbf{B}_0|$ \citep{Cho2003,Zhao2024a}. The orthonormal basis vectors are given by $\hat{\mathbf{e}}_{\parallel}=\hat{\mathbf{b}}_0$, $\hat{\mathbf{e}}_{k\perp} = \hat{\mathbf{b}}_0\times(\hat{\mathbf{k}}\times\hat{\mathbf{b}}_0) / |\hat{\mathbf{b}}_0\times(\hat{\mathbf{k}}\times\hat{\mathbf{b}}_0)|$, and $\hat{\mathbf{e}}_{A} = \hat{\mathbf{e}}_{k\perp}\times\hat{\mathbf{e}}_{\parallel}$. The unit wavevector $\hat{\mathbf{k}} (t,f_{sc})$ was obtained using singular value decomposition (SVD) based on the linearized Gauss’s law for magnetism \citep{Santolik2003}, and the local mean magnetic field was estimated as $\mathbf{B}_0=(\mathbf{B}(t-\tau/2)+\mathbf{B}(t+\tau/2))/2$. Here, $t$ is time, $f_{sc}$ is the spacecraft-frame frequency, and the timescale is $\tau=1/f_{sc}$.

First, magnetic fluctuations were transformed using the Morlet-wavelet transform \citep{Grinsted2004}, yielding Fourier-space fluctuations $\mathbf{\delta B}(t,f_{sc})=[\delta B_{X},\delta B_{Y},\delta B_{Z}]$ in geocentric-solar-ecliptic (GSE) coordinates. The trace magnetic power was calculated as $P(t,f_{sc}) = \delta B_{X}\delta B_{X}^* + \delta B_{Y}\delta B_{Y}^* + \delta B_{Z}\delta B_{Z}^*$. The interval was divided into overlapping windows of $t_{\mathrm{win}} = 30$ min with a 1-min shift for adequate sampling, where $t_{\mathrm{win}}$ is the window duration. To minimize edge effects from finite time series, each transform was performed over a $2t_{\mathrm{win}}$ window, with only the central $t_{\rm win}$ retained for analysis.

Second, based on the linearized induction equation, Alfv\'enic magnetic fluctuations ($\delta \mathbf{B}_{A}$) align with $\hat{\mathbf{e}}_A$, whereas compressible magnetic fluctuations ($\delta \mathbf{B}_{C}$) align with $\hat{\mathbf{e}}_{C}=\hat{\mathbf{e}}_{A}\times\hat{\mathbf{k}}$ and lie in the $\hat{k}\hat{b}_0$ plane (Fig. \ref{fig:1}). The corresponding components were calculated as
 \begin{eqnarray}
    \delta B_{A} = \mathbf{\delta B} \cdot \hat{\mathbf{e}}_A,\\
    \delta B_{C} = \mathbf{\delta B} \cdot \hat{\mathbf{e}}_C,
 \end{eqnarray}
with the associated power,
\begin{eqnarray}
   P_{A}(t,f_{sc}) = \delta B_{A}\delta B_{A}^*,\\
   P_{C}(t,f_{sc}) = \delta B_{C}\delta B_{C}^*.
\end{eqnarray}

Third, we calculated two key angles: (1) $\eta$, the angle between $\mathbf{\delta B}$ and the $\hat{k}\hat{b}_0$ plane, defined as $\eta=arctan(\sqrt{P_{A}/P_{C}})$ to eliminate phase-difference effects between $\delta B_A$ and $\delta B_C$. As shown in Fig. \ref{fig:1}, $\eta$ is a key parameter that characterizes the polarization of magnetic fluctuations. For $\eta = 90^\circ$, the fluctuations are perpendicular to the $\hat{k}\hat{b}_0$ plane and are thus incompressible (Alfvénic). For $\eta = 0^\circ$, the fluctuations lie within the $\hat{k}\hat{b}_0$ plane, corresponding to compressible modes. Intermediate values of $\eta$ indicate a superposition of Alfvénic and compressible fluctuations. (2) $\theta$, the angle between $\hat{\mathbf{k}}$ and $\hat{\mathbf{b}}_0$. Owing to the antiparallel ambiguity of the SVD method, $\theta$ was restricted to $[0^\circ, 90^\circ]$, with values $>90^\circ$ remapped to $\theta =180^\circ - \theta$. To minimize uncertainty in defining the $\hat{k}\hat{b}_0$ plane, cases with $\eta<5^\circ$ or $\theta<5^\circ$ were excluded. This has no impact on the main results, as their behavior is broadly similar to that at larger angles (Fig. \ref{fig:2}). Further details on the sensitivity of the results to the $\theta$ and $\eta$ thresholds are provided in Figs. \ref{fig:5} and \ref{fig:6} of Appendix A.

Fourth, we determined wavevectors using multi-spacecraft timing analysis based on phase differences, restricted to intervals with high tetrahedron quality factor ($TQF>0.8$) \citep{Pincon2008}; see \cite{Zhao2024a} for details. Unlike SVD method, which only provides the best estimate of unit wavevector ($\mathbf{\hat{k}}$) representing the normal direction of the overall $\delta\mathbf{B}$ plane formed by the superposition of multiple modes, the timing analysis yields the actual wavevector $\mathbf{k}_{M}$ ($M=A$ for Alfv\'enic, $M=C$ for compressible) from phase differences of $\delta B_A$ and $\delta B_C$, respectively. Due to the superposition of different modes, each with its own dispersion relation and distinct wavevectors at the same frequency, $\mathbf{k}_{M}$ is not necessarily aligned with the SVD-derived $\mathbf{\hat{k}}$. The timing analysis is considered valid only when $\mathbf{k}_M$ lies within the $\hat{k}\hat{b}_0$ plane defined by $\hat{\mathbf{k}}$, ensuring consistency in the $\hat{k}\hat{b}_0$ coordinates during mode decomposition, even if the two vectors differ within that plane. Accordingly, the wavenumber analysis is restricted to fluctuations with a small angle $\xi\leq20^\circ$ between $\mathbf{k}_M$ and the $\hat{k}\hat{b}_0$ plane (Fig. \ref{fig:4}(b)), to ensure reliable wavevector estimates. The main results are insensitive to the choice of $\xi$ threshold ($\xi<10^\circ$, $15^\circ$, and $20^\circ$ in Fig. \ref{fig:7} of Appendix B). Wavenumbers were further limited to $1/(100d_{sc})<k<{min}(0.1/d_p,\pi/d_{sc})$, with $k_\parallel\sim\mathbf{k}_{M}\cdot\mathbf{\hat{b}}_0$ and $k_\perp=\sqrt{k^2-k_\parallel^2}$, where $d_{sc}$ is the spacecraft separation and $d_p$ is the proton inertial length. These restrictions apply only to the wavenumber analysis and not to the angle analysis (Figs. \ref{fig:2}, \ref{fig:3}, \ref{fig:4}(a)). This approach enables direct retrieval of energy spectra in wavenumber space from frequency space, independent of spatiotemporal assumptions such as the Taylor hypothesis \citep{Taylor1938}.


Finally, the data were divided into $45$ equal-width bins in $\theta$ with width $\Delta\theta=2^\circ$, $90$ equal-width bins in $\eta$ with $\Delta\eta=1^\circ$, and $50$ logarithmically spaced bins in $k_\parallel$ with $\Delta k_\parallel$, to construct $\theta-\eta-k_\parallel$ distributions of magnetic energy. For each $\theta$, $\eta$, or $k_\parallel$ bin, the trace, Alfvénic, and compressible magnetic energy were estimated as $D =\frac{\iint P(t,f_{sc})dtdf_{sc}}{\Delta x}$, $D_A=\frac{\iint P_{A}(t,f_{sc})dtdf_{sc}}{\Delta x}$, and $D_C=\frac{\iint P_{C}(t,f_{sc})dtdf_{sc}}{\Delta x}$, where $\Delta x=\Delta\theta$, $\Delta\eta$, or $\Delta k_\parallel$. To capture inertial-range MHD fluctuations, we restricted to $4/t_{\rm{win}}<f_{sc}<0.05$ Hz. The lower cutoff ensures sufficient averaging over multiple oscillations for statistical robustness and avoids contamination from the energy-containing range, and the upper limit corresponds to half of the typical proton gyrofrequency ($\sim0.1$ Hz) in the solar wind.


\section{Observations}

We analyzed six solar wind intervals with varying plasma conditions: three in slow wind ($\langle V_p\rangle<450$ km/s) and three in fast wind ($\langle V_p\rangle>450$ km/s), where $\langle V_p\rangle$ is effectively the solar wind speed, as the spacecraft was nearly stationary with respect to the flow. All intervals were in the pristine solar wind, confirmed by spectrograms of ion differential energy fluxes showing no reflected ions from Earth's bow shock. The turbulence was fully developed, with magnetic spectral slopes close to $-5/3$ or $-3/2$ over $f_{sc}=0.001-0.05$ Hz. To ensure the reliability of the linear decomposition, the relative fluctuation amplitude $\delta B/\langle B\rangle$ is required to be less than 0.4. Solar wind turbulence is intrinsically compressible, containing a nonnegligible fraction ($25.2\%\pm2.8\%$) of compressible magnetosonic modes derived from the six intervals, with the fraction systematically decreasing as $\beta$ increases. The key parameters are summarized in Table \ref{tab:table1}.

\begin{figure*}[ht!]
\centering
\includegraphics[scale=0.26]{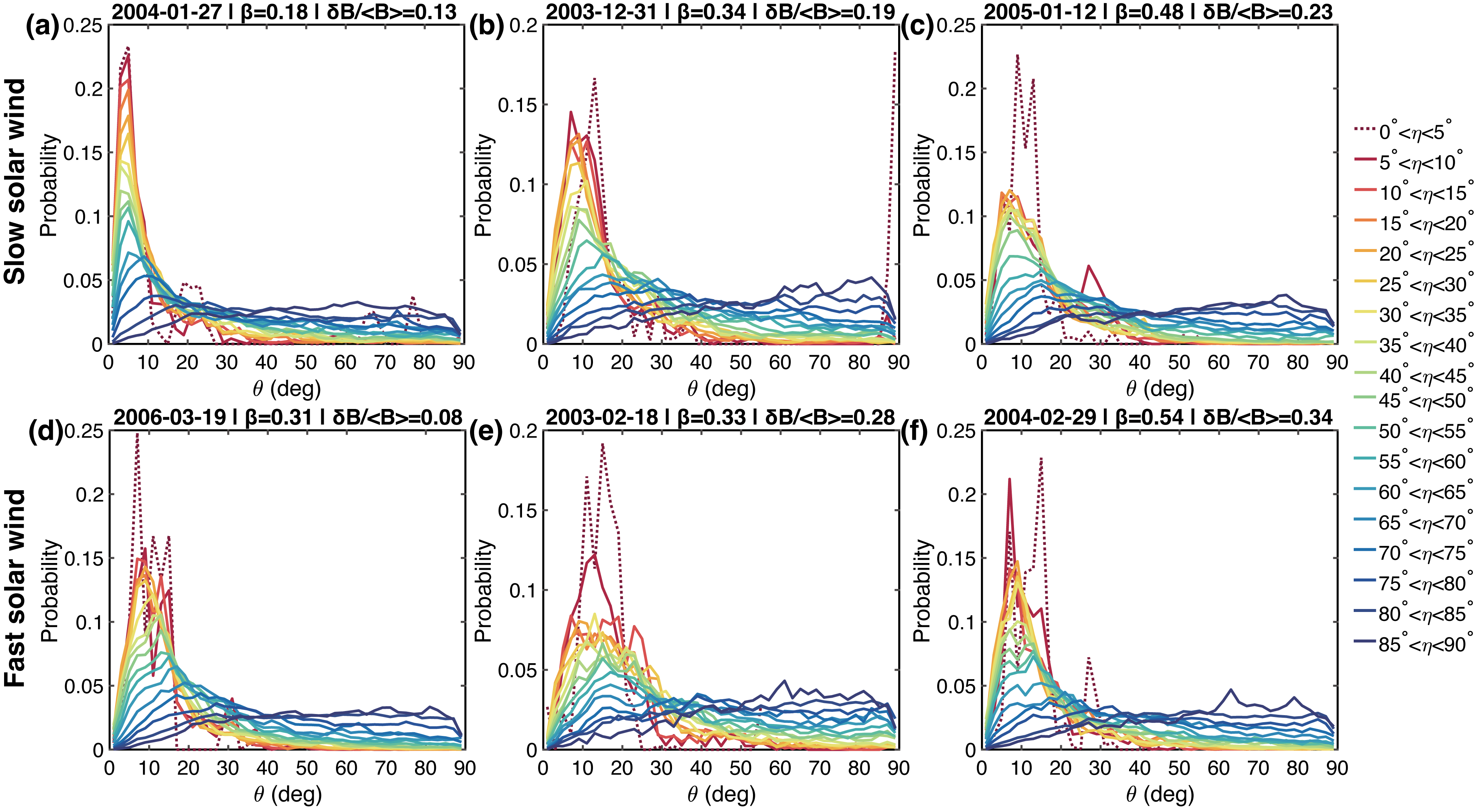}
\caption{\label{fig:2} Probability distributions of the angle $\theta$ in slow solar wind (a-c) and fast solar wind (d-f). Colors denote $\eta$ ranges.}
\end{figure*}

Fig. \ref{fig:2} presents the probability distributions of $\theta$ for different $\eta$ ranges in slow and fast solar wind. Each panel is labeled with the corresponding $\beta$ and $\delta B/\langle B\rangle$, where $\langle\cdot\rangle$ denotes the interval average. The probability decreases monotonically with increasing $\eta$ at $\theta<30^\circ$, whereas for $\theta>30^\circ$ it increases with $\eta$. This reveals a universal magnetic geometry: quasi-parallel (slab) fluctuations at $\theta<30^\circ$ are preferentially associated with fluctuations confined to the $\hat{k}\hat{b}_0$ plane, i.e., more compressible. In contrast, at $\theta>30^\circ$, fluctuations deviate more strongly from, or approach perpendicularity to, the $\hat{k}\hat{b}_0$ plane. This monotonic $\theta-\eta$ dependence is robust across all intervals and is independent of plasma and field parameters, such as $\langle V_p\rangle$ and $\delta B/\langle B\rangle$, in low-$\beta$ plasmas.

\begin{figure*}[ht!]
\centering
\includegraphics[scale=0.28]{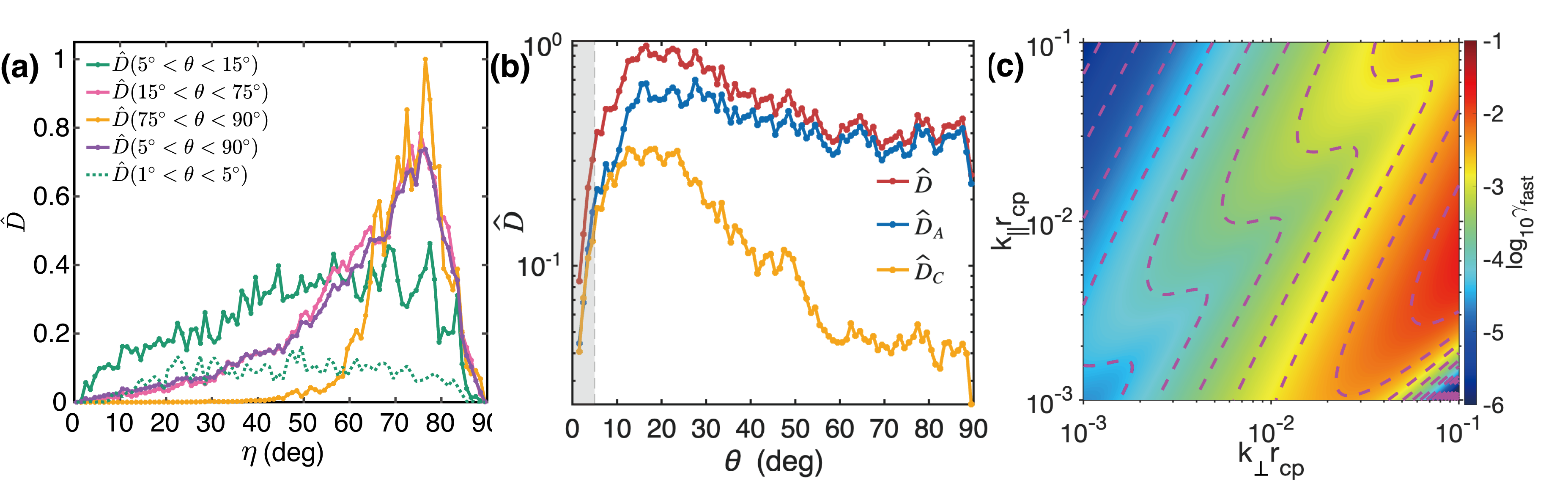}
\caption{\label{fig:3} Results for the 18 February 2003 interval. (a) Normalized trace magnetic energy ($\hat{D}$) for five $\theta$ ranges. (b) $\theta$ distributions of $\hat{D}$, Alfv\'enic ($\hat{D}_A$), and compressible ($\hat{D}_C$) magnetic energy. Gray-shaded regions with $\theta<5^\circ$ are excluded from analysis due to large uncertainties in defining the $\hat{k}\hat{b}_0$ plane. (c) Theoretical fast-mode damping rate ($\gamma_{\rm fast}$), shown as color contours with dashed curves. $k_\parallel$ and $k_\perp$ are normalized by the proton gyro-radius ($r_{cp}$).}
\end{figure*}

We further examine the relationship between magnetic anisotropy and mode composition in compressible MHD turbulence. Fig. \ref{fig:3} presents results from a representative interval on 18 February 2003 (additional events are shown in Figs. \ref{fig:8} and \ref{fig:9} of Appendices C and D). The magnetic energy $\hat{D}=D/D_{max}$ was normalized by its maximum across all $\eta$ and $\theta$ ranges. In Fig. \ref{fig:3}(a), quasi-parallel (slab) energy with $5^\circ < \theta < 15^\circ$ (green) exhibits only weak $\eta$ dependence, with energy broadly distributed. The green dashed curve for $1^\circ<\theta<5^\circ$ follows a similar trend, albeit with larger uncertainty in defining the $\hat{k}\hat{b}_0$ plane. In contrast, quasi-perpendicular (2D) energy with $75^\circ < \theta < 90^\circ$ (yellow) is predominantly concentrated at $\eta>60^\circ$, indicating magnetic fluctuations nearly orthogonal to both $\hat{\mathbf{k}}$ and $\hat{\mathbf{b}}_0$. Intermediate-$\theta$ energy with $15^\circ < \theta < 75^\circ$ (pink) exhibits transitional behavior and closely tracks the overall energy distribution with $5^\circ < \theta < 90^\circ$ (purple).


Fig. \ref{fig:3}(b) shows the $\theta$ distribution of magnetic energy, where $\hat{D}$, $\hat{D}_A$, and $\hat{D}_C$ denote the normalized trace, Alfvénic, and compressible components, respectively. The magnetic fluctuations are composed of Alfvén and compressible modes. $\hat{D}_A$ spans over a broad range of $\theta$, showing only a slight preference for small $\theta$ and approaching $\hat{D}$ at large $\theta$. In contrast, $\hat{D}_C$ is predominantly concentrated in the quasi-parallel direction with $\theta<30^\circ$ and decreases steadily for $\theta>20^\circ$. This trend highlights an enhanced contribution of compressible fluctuations in the slab component and a dominant role of Alfvénic fluctuations in the quasi-2D component.

Distinct from Alfvén modes, compressible modes undergo strong transit-time damping, a magnetic-mirror–driven Landau-type process in collisionless plasmas \citep{ginzburg1962,Yan2004}. In low-$\beta$ plasmas, compressible magnetic energy is dominated by fast modes \citep{Zhao2021,Galtier2023}; thus, the analysis reduces to how fast-mode damping modulates their otherwise isotropic cascade. The theoretical damping rate of fast modes in low-$\beta$ limits is given by \citep{ginzburg1962,Yan2004}
\begin{eqnarray}
\gamma_{\rm fast}=\frac{\sqrt{\pi\beta}}{4}\omega\frac{sin^2\theta}{cos\theta}[\sqrt{\frac{m_e}{m_p}}exp(-\frac{m_e}{m_p\beta cos^2\theta}) \\ \nonumber+ 5exp(-\frac{1}{\beta cos^2\theta})],
\label{eq:5}
\end{eqnarray}
where $\omega$ is the wave frequency, and $m_p$ and $m_e$ are proton and electron masses. Fig. \ref{fig:3}(c) shows $\gamma_{\rm fast}$ for the 18 February 2003 interval, in excellent agreement with numerical solutions from the dispersion relation solver WHAMP (Waves in Homogeneous Anisotropic Multicomponent Plasma) \citep{Rönnmark2024} (Fig. \ref{fig:10} in Appendix E). The damping rate is negligible for quasi-parallel propagation but becomes pronounced toward quasi-perpendicular directions, consistent with Fig. \ref{fig:3}(b), which shows compressible energy is concentrated at $\theta < 30^\circ$ and nearly absent at $\theta > 60^\circ$.

\begin{figure*}[ht!]
\centering
\includegraphics[scale=0.23]{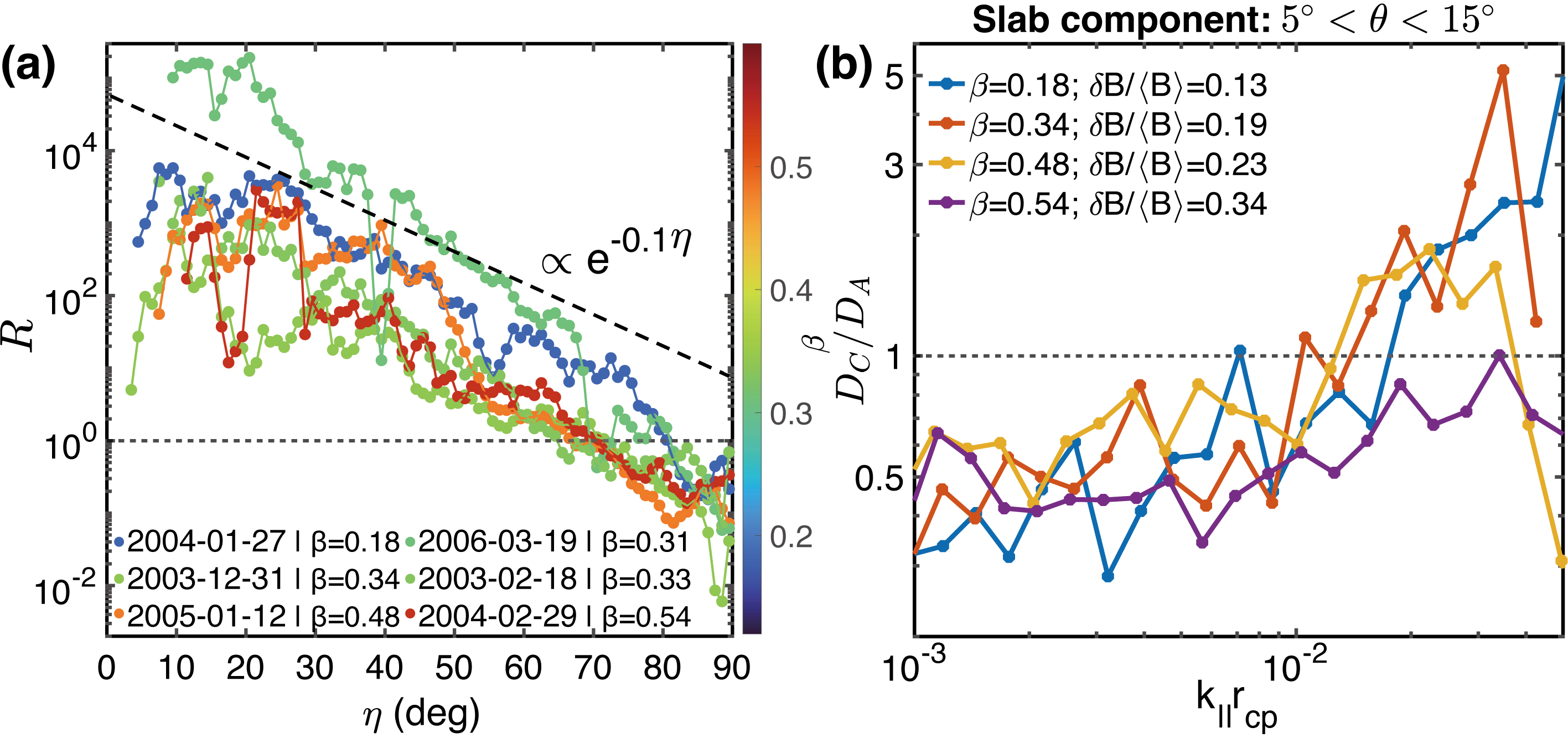}
\caption{\label{fig:4} (a) $\eta$-dependence of magnetic anisotropy ($R(\eta)\equiv\frac{\hat{D}(5^\circ < \theta < 15^\circ)}{\hat{D}(75^\circ < \theta < 90^\circ)}$) for all intervals. Colors denote $\beta$ values. Dashed line represents $R\propto e^{-0.1\eta}$. (b) Ratio of $D_C/D_A$ for quasi-parallel (slab) component with $5^\circ<\theta<15^\circ$ as a function of $k_\parallel r_{cp}$ at $\xi<20^\circ$.} 
\end{figure*}

Fig. \ref{fig:4}(a) shows $\eta$ distributions of magnetic anisotropy, defined as $R(\eta)\equiv\frac{\hat{D}(5^\circ < \theta < 15^\circ)}{\hat{D}(75^\circ < \theta < 90^\circ)}$. Across all intervals, $R$ decreases exponentially with $\eta$, with a universal decay coefficient of $\sim0.1$ (see fits in Fig. \ref{fig:11} of Appendix F). This means that as the oscillation direction of $\delta \mathbf{B}$ departs further from the $\hat{k}\hat{b}_0$ plane, the ratio of parallel to perpendicular energy decreases monotonically, and thus magnetic topology becomes increasingly quasi-2D. Additionally, in the quasi-perpendicular case, $D_C/D_A$ remains very small (Fig. \ref{fig:3}(b)), independent of the magnitude of $k_\perp$. For quasi-parallel (slab) component with $5^\circ<\theta<15^\circ$, Fig. \ref{fig:4}(b) shows the ratio $D_C/D_A$ as a function of $k_\parallel r_{cp}$. For $k_\parallel r_{cp}<10^{-2}$, $D_C/D_A$ remains nearly constant at $\sim1/2$ across all intervals. For $k_\parallel r_{cp}>10^{-2}$, $D_C/D_A$ increases with $k_\parallel$, indicating a growing compressible contribution and enhanced anisotropy at smaller scales, consistent with the decrease of $\gamma_{fast}$ with $k_\parallel$ in Fig. \ref{fig:3}(c). In low-$\beta$ limit ($\beta\rightarrow 0$), $D_C/D_A$ exceeds unity, suggesting that compressible modes can dominate. At relatively higher $\beta$ (purple), the compressible contribution stays limited, consistent with a stronger fast-mode damping as $\beta$ approaches unity \citep{Yan2004}. A more quantitative assessment of this dependence requires a larger statistical sample, which is beyond the scope of this study but will be pursued in future work.

 \section{Discussion}

In this study, small-amplitude fluctuations are decomposed into Alfvénic and compressive modes based on their polarization characteristics rather than linear dispersion relations; therefore, they should not be interpreted as a simple superposition of linear MHD eigenmodes. Although this approach cannot entirely eliminate contributions from nonpropagating or structure modes, which \cite{Zank2023} notes are non-negligible, their impact on our results is expected to be minor. The entropy mode, which produces no magnetic fluctuations, does not affect magnetic anisotropy. Magnetic-island (flux-rope) modes, characterized by transverse magnetic fields with $\delta \mathbf{B}_\perp \neq0$ and $\delta \mathbf{B}_\parallel =0$, are largely excluded from the wavenumber analysis by restricting the plasma-frame frequency to $f_{rest} > 4/t_{win}$, where $f_{rest} = f_{sc}-\mathbf{k}_M\cdot \mathbf{V}_p/(2\pi)$ via the Doppler shift. In the angular analysis, magnetic-island modes may still be partially included in either $\delta B_A$ or the perpendicular component of $\delta B_C$. Nevertheless, our main conclusion that compressible modes contribute significantly to the slab component at small scales (large $k$) is unaffected, because nonpropagating magnetic-island structures primarily influence the large-scale (small-$k$) part of the energy spectrum.

\section{Conclusion}


We present Cluster observations demonstrating that mode composition plays a central role in shaping magnetic anisotropy in compressible MHD turbulence within the low-$\beta$ solar wind. The principal findings are summarized below.

\begin{enumerate}
    \item Solar wind turbulence is intrinsically compressible, with a nonnegligible fraction ($25.2\%\pm2.8\%$) of compressible magnetosonic modes, particularly fast modes in low-$\beta$ plasmas, and this fraction systematically decreases with increasing $\beta$.
    \item Quasi-parallel (slab) energy at $\theta<30^\circ$ is predominantly associated with compressible fluctuations confined to the $\hat{k}\hat{b}_0$ plane, whereas at $\theta>30^\circ$ the fluctuations deviate markedly from, or approach perpendicularity (Fig. \ref{fig:2}).
    \item Quasi-perpendicular (2D) energy is almost purely Alfvénic, whereas quasi-parallel (slab) energy contains both Alfvénic and compressible modes, a feature closely linked to the collisionless damping of compressible modes at high obliquity (Fig. \ref{fig:3}). 
    \item Magnetic anisotropy decreases exponentially with $\eta$, with a universal coefficient of $\sim0.1$ (Fig. \ref{fig:4}).
\end{enumerate}

Our observations refine the traditional `slab+2D' picture, which emphasized Alfvénic modes, by revealing that both Alfvén and compressible (magnetosonic) modes contribute significantly. Alfvénic fluctuations are broadly distributed in propagation angle, whereas compressible fluctuations are concentrated near the quasi-parallel (slab) direction (Fig. \ref{fig:3}(b)). Notably, for $\beta\rightarrow0$, compressible modes become dominant within the slab component at small scales ($k_\parallel r_{cp}>10^{-2}$; Fig. \ref{fig:4}(b)). Since the plasma parameters are generic, the universal magnetic geometry and mode-dependent anisotropy are likely to occur across diverse plasma environments. Future work should test this universality in more dynamic settings, such as planetary shocks, magnetosheaths, and reconnection regions. These findings have implications that extend beyond turbulence, encompassing particle transport, acceleration, and magnetic reconnection.

\begin{acknowledgments}
We thank Gary Zank and Xiangrong Fu for valuable suggestions
and discussions. We also acknowledge the members of the Cluster spacecraft team and NASA’s Coordinated Data Analysis Web. The Cluster data and OMNI data are available at \url{https://cdaweb.gsfc.nasa.gov}. Data analysis was performed using the IRFU-MATLAB analysis package \citep{Khotyaintsev2024} available at \url{https://github.com/irfu/irfu-matlab}.
\end{acknowledgments}

\begin{contribution}
H.Y. initiated and designed the project. S.Z. and T.Z.L. developed and implemented the data processing methods. S.Z. performed the observation data processing. S.Z., H.Y., and T.Z.L. contributed to the theoretical analysis of the main results. All authors contributed to the writing, editing, and approval of the manuscript. 
\end{contribution}

\appendix

\section{Sensitivity of the results to the $\theta$ and $\eta$ thresholds}

\begin{figure*}[ht!]
\centering
\includegraphics[scale=0.17]{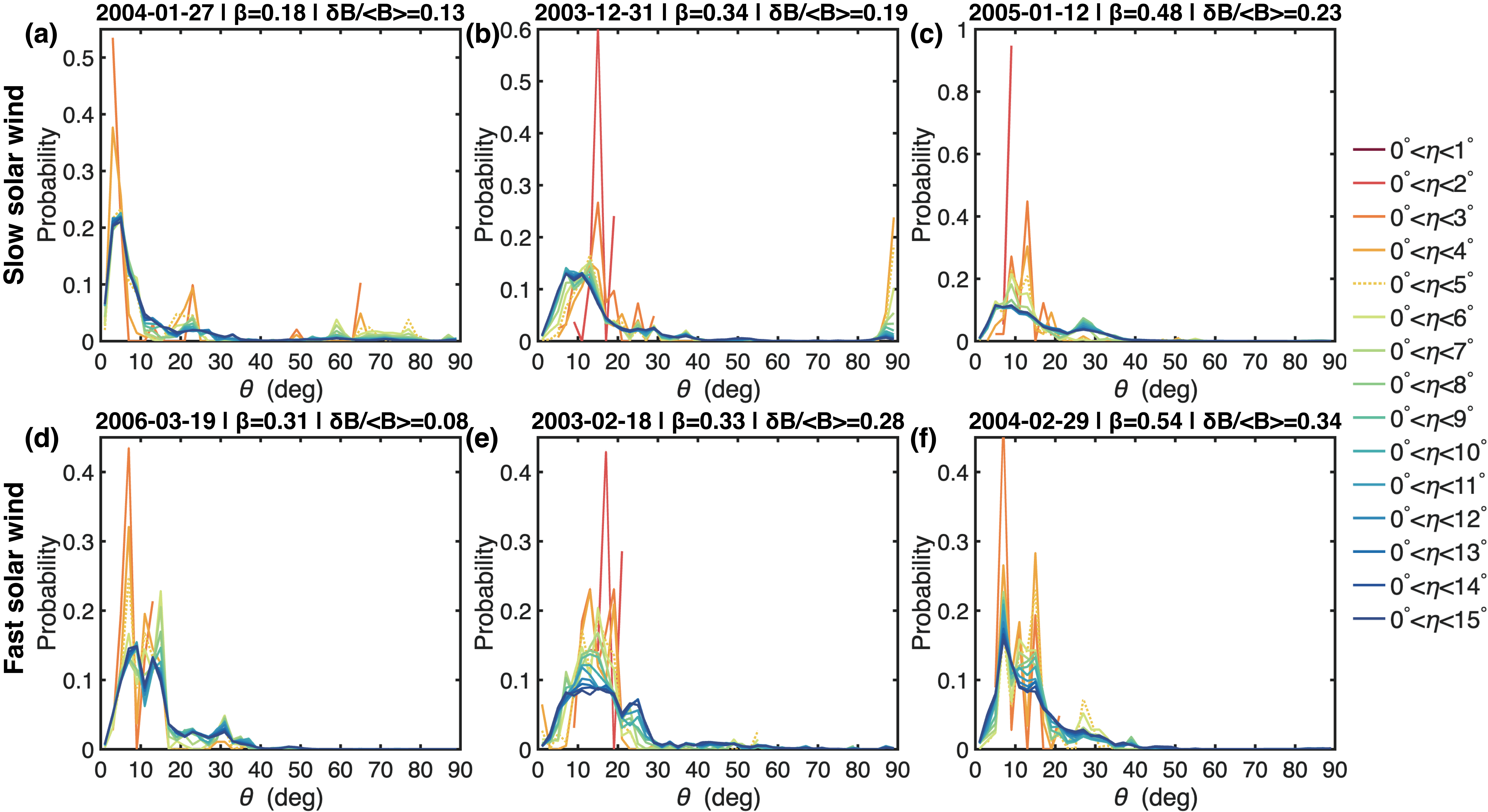}
\caption{\label{fig:5} Probability distributions of the angle $\theta$ in slow solar wind (a-c) and fast solar wind (d-f). Colors denote $\eta$ ranges.}
\end{figure*}

\begin{figure*}[ht!]
\centering
\includegraphics[scale=0.17]{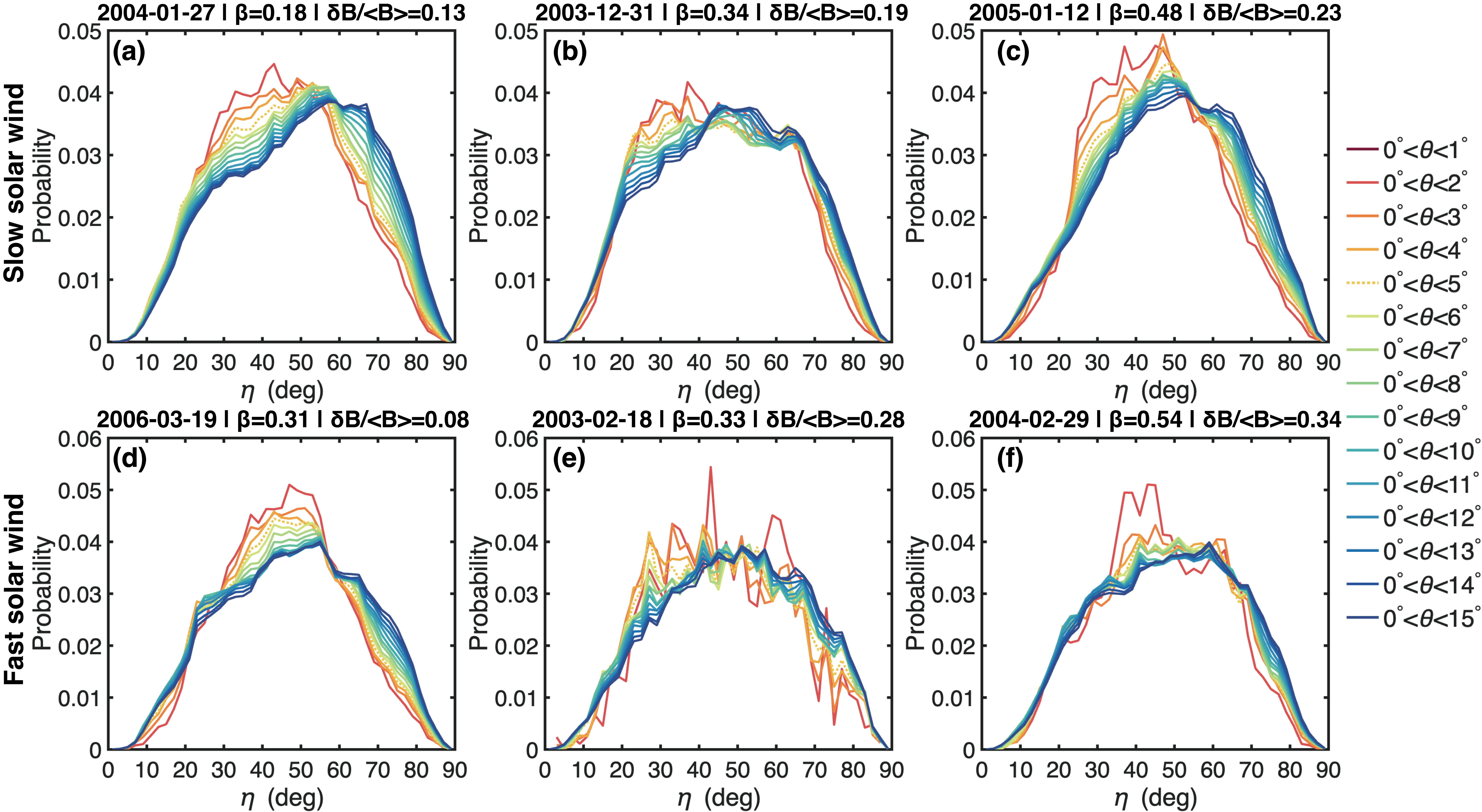}
\caption{\label{fig:6} Probability distributions of the angle $\eta$ in slow solar wind (a-c) and fast solar wind (d-f). Colors denote $\theta$ ranges.}
\end{figure*}

Fig. \ref{fig:5} presents the probability distributions of $\theta$ in the slow and fast solar wind, where the colors correspond to $\eta$ ranges of $0^\circ-1^\circ$, $0^\circ-2^\circ$, $0^\circ-3^\circ$, and so on up to $0^\circ-15^\circ$. Here, $\theta$ is the angle between $\hat{\mathbf{k}}$ and $\hat{\mathbf{b}}_0$, and $\eta$ is the angle between $\mathbf{\delta B}$ and the $\hat{k}\hat{b}_0$ plane. The distributions vary significantly with $\eta$ when $\eta<5^\circ$ but become progressively more stable as $\eta$ increases beyond $5^\circ$. Fig. \ref{fig:6} shows the probability distributions of $\eta$, where the colors correspond to $\theta$ ranges of $0^\circ-1^\circ$, $0^\circ-2^\circ$, $0^\circ-3^\circ$, and so on up to $0^\circ-15^\circ$. The distributions vary significantly for $\theta < 5^\circ$ but stabilize for $\theta > 5^\circ$, likely because the uncertainty in the coordinates constructed from $\hat{\mathbf{k}}$ and $\hat{\mathbf{b}}_0$ increases as $\theta$ approaches $0^\circ$. Therefore, we adopt $\eta>5^\circ$ and $\theta>5^\circ$ as the thresholds in this study.

\section{Sensitivity of the results to the $\xi$ threshold for the wavenumber analysis}

\begin{figure*}[ht!]
\centering
\includegraphics[scale=0.17]{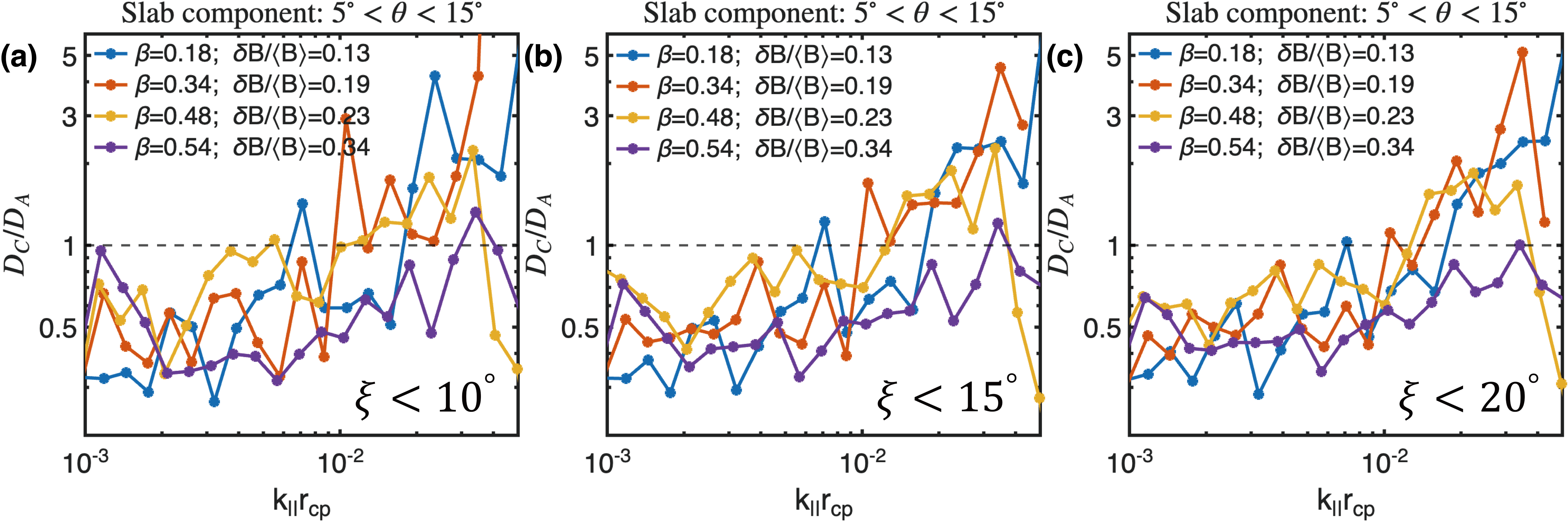}
\caption{\label{fig:7} Ratio of $D_C/D_A$ for quasi-parallel (slab) component with $5^\circ<\theta<15^\circ$ as a function of $k_\parallel r_{cp}$ when angle $\xi<10^\circ$, $15^\circ$, and $20^\circ$.}
\end{figure*}

Fig. \ref{fig:7} shows the ratio $D_C/D_A$ as a function of $k_\parallel r_{cp}$ for $\xi<10^\circ$, $15^\circ$, and $20^\circ$, where $\xi$ is the angle between the timing-derived $\mathbf{k}_M$ and the $\hat{k}\hat{b}_0$ plane defined by the SVD-derived $\hat{\mathbf{k}}$. All three cases exhibit the same overall trend. For $k_\parallel r_{cp}<10^{-2}$, $D_C/D_A$ remains nearly constant at $\sim1/2$, whereas for $k_\parallel r_{cp}>10^{-2}$, $D_C/D_A$ increases with $k_\parallel$. In the low-$\beta$ limit ($\beta\rightarrow 0$), $D_C/D_A$ exceeds unity, whereas at higher $\beta$ (purple), the compressible contribution remains limited. Increasing $\xi$ includes more data points, producing smoother but less accurate wavenumber estimates. Nevertheless, the main results are insensitive to the choice of $\xi$ threshold.

\section{$\eta$-dependence of magnetic energy}

\begin{figure*}[ht!]
\centering
\includegraphics[scale=0.26]{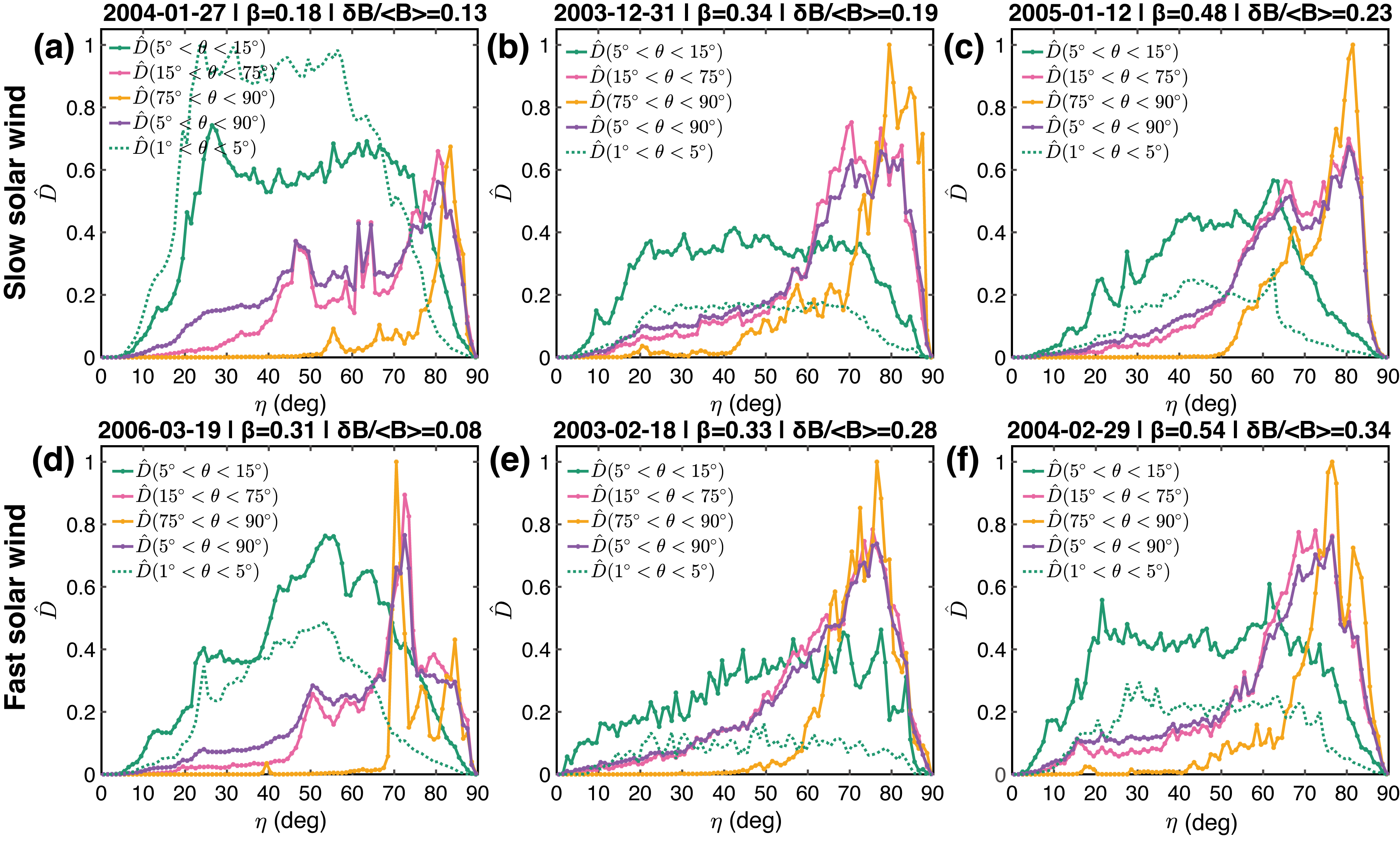}
\caption{\label{fig:8} $\eta-$dependence of trace magnetic energy ($\hat{D}$) in \textbf{slow} solar wind (a-c) and \textbf{fast} solar wind (d-f).}
\end{figure*}

Fig. \ref{fig:8} shows the $\eta$-dependence of magnetic energy in slow and fast solar wind, where $\eta$ is the angle between magnetic fluctuations $\mathbf{\delta B}$ and the $\hat{k}\hat{b}_0$ plane. Across all intervals, the $\eta$ distributions of magnetic energy exhibit the same overall trend, independent of solar wind speed, $\beta$, or fluctuation amplitude. Quasi-parallel energy with $5^\circ < \theta < 15^\circ$ (green) exhibits only a weak dependence on $\eta$, with energy distributed broadly. In contrast, quasi-perpendicular energy with $75^\circ < \theta < 90^\circ$ (yellow) is concentrated primarily at $\eta>60^\circ$, indicating that the associated magnetic fluctuations are nearly orthogonal to both $\hat{\mathbf{k}}$ and $\hat{\mathbf{b}}_0$. Intermediate-$\theta$ energy with $15^\circ < \theta < 75^\circ$ (pink) displays transitional behavior, closely following the overall energy distribution with $5^\circ < \theta < 90^\circ$ (purple). 

\section{$\theta$-dependence of magnetic energy}

\begin{figure*}[ht!]
\centering
\includegraphics[scale=0.27]{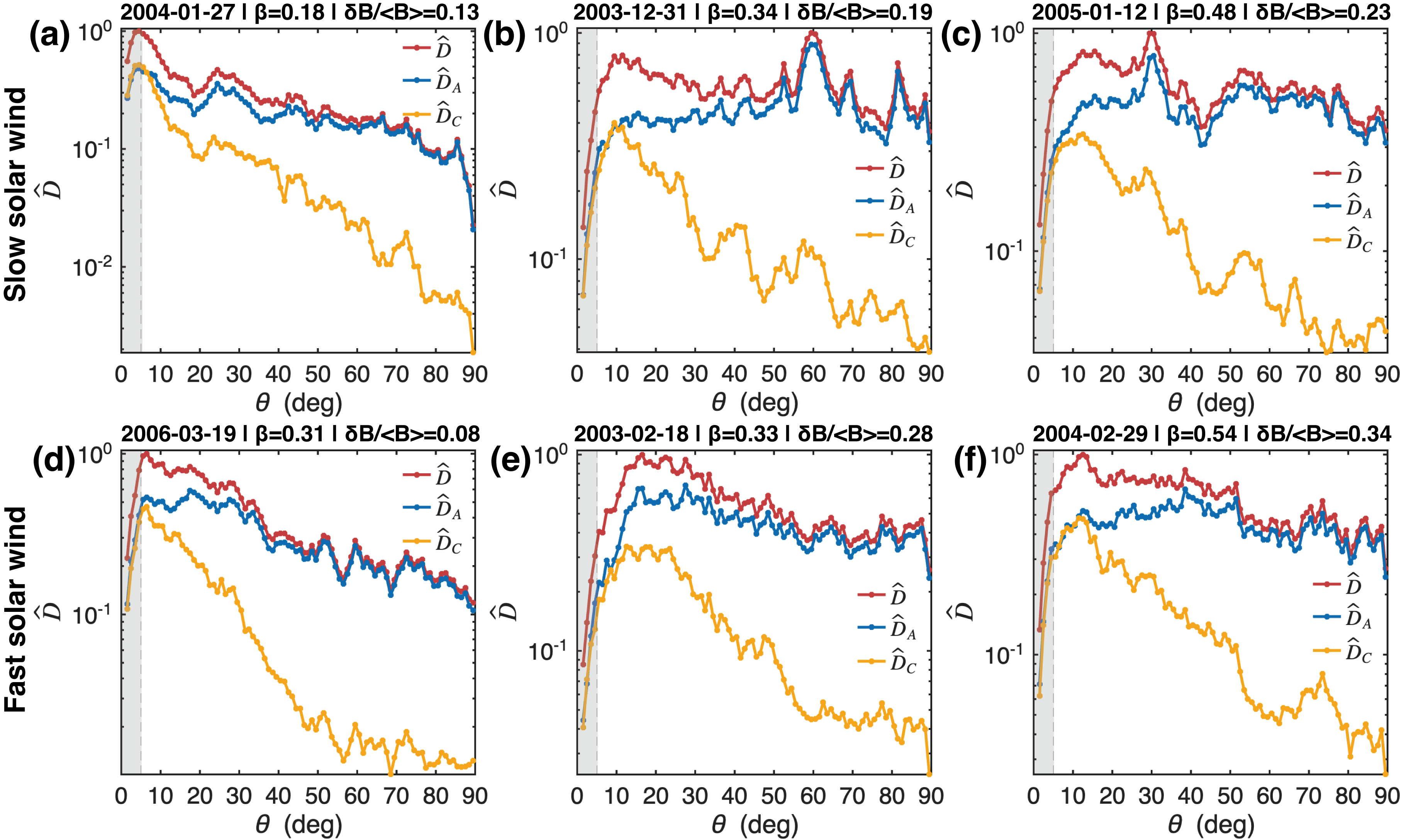}
\caption{\label{fig:9} $\theta$-dependence of $\hat{D}$, $\hat{D}_A$, and $\hat{D}_C$ for the six solar wind intervals. Gray-shaded regions ($\theta<5^\circ$) are excluded from analysis due to large uncertainties in defining the $\hat{k}\hat{b}_0$ plane.}
\end{figure*}

Fig. \ref{fig:9} shows the $\theta$-dependence of magnetic energy for six solar wind intervals, where $\theta$ is the angle between $\hat{\mathbf{k}}$ and $\hat{\mathbf{b}}_0$. Across all events, $\hat{D}_A$ dominates, especially at $\theta>30^\circ$, and closely tracks the trace energy $\hat{D}$, indicating predominantly Alfvénic magnetic fluctuations. In contrast, $\hat{D}_C$ is mainly concentrated at $\theta<30^\circ$. In the quasi-parallel regime, $\hat{D}_C$ can even rival $\hat{D}_A$, but decreases steadily with increasing $\theta$. This trend demonstrates the enhanced role of compressible fluctuations at small propagation angles.

\section{Fast-mode damping rate from WHAMP}

\begin{figure*}[ht!]
\centering
\includegraphics[scale=0.21]{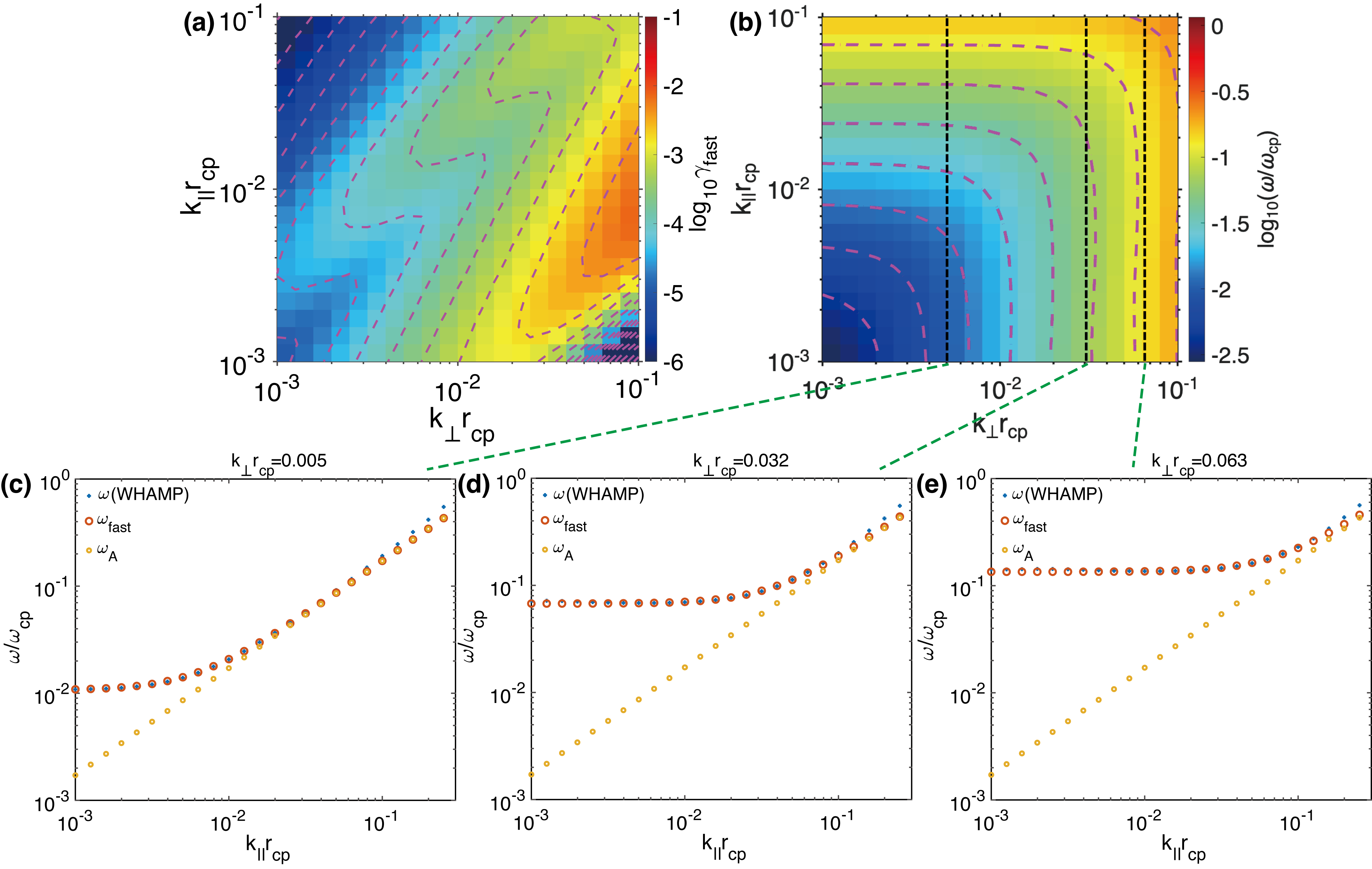}
\caption{\label{fig:10} WHAMP results using parameters from the 18 February 2003 interval. (a) Fast-mode damping rate ($\gamma_{\rm fast}$) from WHAMP. (b) $k_\parallel-k_\perp$ distributions of frequencies. (c-e) Dispersion relations at $k_\perp r_{cp}=0.005$, $0.032$, and $0.063$, respectively. }
\end{figure*}

Fig. \ref{fig:10}(a) shows the damping rate of fast modes calculated using WHAMP (Waves in Homogeneous Anisotropic Multicomponent Plasma), in good agreement with the theoretical fast-mode damping rate in Fig. \ref{fig:3}(c) of the main text. The numerical solution was identified as the fast modes based on three criteria. (1) Isotropic frequency distribution in $k_\parallel-k_\perp$ spectra (Fig. \ref{fig:10}(b)). (2) Right-hand polarization, quantified by $S_3=\Im(E_{\perp1}E_{\perp2}^*)<0$, where $E_{\perp1}$ and $E_{\perp2}$ are the two components of electric fields perpendicular to the mean magnetic field. (3) Consistency with the dispersion relations of fast modes in Fig. \ref{fig:10}(c-e).

\section{$\eta$-dependence of magnetic anisotropy}

\begin{figure*}[ht!]
\centering
\includegraphics[scale=0.25]{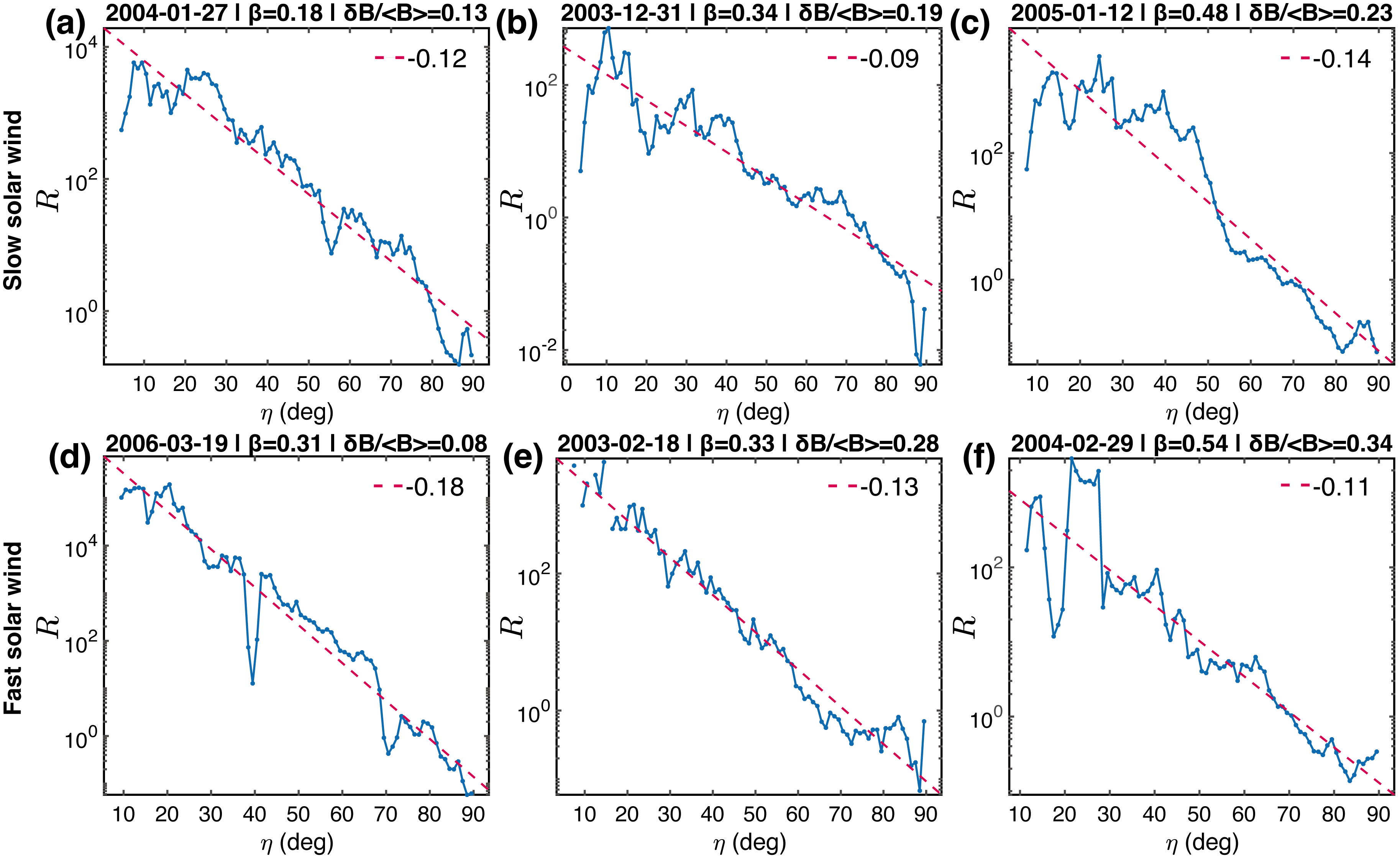}
\caption{\label{fig:11} $\eta$-dependence of magnetic anisotropy ($R$) for the six solar wind intervals. The red dashed lines represent log-linear slopes (exponential decay rates), indicated in the upper-right corner of each panel.}
\end{figure*}

Fig. \ref{fig:11} shows the $\eta$-dependence of magnetic anisotropy, defined as $R(\eta)=\frac{\hat{D}(5^\circ < \theta < 15^\circ)}{\hat{D}(75^\circ < \theta < 90^\circ)}$. In all events, the anisotropy decreases approximately exponentially with $\eta$, with a universal decay coefficient of $\sim0.1$.

\setlength{\tabcolsep}{2pt}
\begin{table*}[ht!]
\caption{\label{tab:table1} Magnetic field and plasma parameters in pristine solar wind.}
\begin{ruledtabular}
\begin{tabular}{ccccccccccccccccccc}
 No. & Date& Start Time & End Time & $\langle|B|\rangle$& $\langle N_p \rangle$& $\langle T_p \rangle$& $\langle V_p\rangle$ & $\phi_{VB}$  &$f_{cp}$&$r_{cp}$&$d_p$&$\beta$ &$\delta B/\langle B\rangle$&$C_{\parallel}$& $\alpha_B$&TQF&$\langle\frac{D_C}{D}\rangle_f$\\ 
 & & (UT)& (UT) & ($nT$) &($cm^{-3}$) &($eV$) & ($km s^{-1}$) & ($deg$) & ($Hz$)&($km$)&($km$)&\\
 \hline
 
1&2004-01-27&00:36:00&01:18:00 &9.7 &6.3 & 6.8&428&79&0.15&27&91&0.18 &0.13&0.06 & 1.7&0.92 &30.2$\%$\\
2&2003-12-31&10:48:00&11:30:00 &11.3&18.9& 5.7&432&74&0.17&22& 52&0.34 &0.19&0.07 &1.6&0.89&23.3$\%$\\
3&2005-01-12&02:00:00&02:42:00 &13.9&27.4& 8.4&439&83&0.21&21& 44&0.48 &0.23&0.09&1.7&0.85 &22.3$\%$ \\
4&2006-03-19&20:34:00&21:16:00 &6.7 & 2.5&14.4&625&56&0.10&58&145&0.31 &0.08&0.03&1.5&0.43&24.6$\%$\\
5&2003-02-18&00:18:00&01:00:00 &15.5& 5.9& 33.0 &668&77&0.24&38& 94&0.33&0.28&0.14&1.5&0.88&24.0$\%$ \\
6&2004-02-29&04:03:00&04:45:00 &9.6& 2.7& 45.3 &650&71&0.15&71& 138&0.54&0.34&0.11&1.5&0.98&26.5$\%$ \\
\end{tabular}
\end{ruledtabular}
\tablenotetext{a}{
$\langle\cdot\rangle$ denotes the average over whole interval ($42$ minutes);
$|B|$ is the magnetic field magnitude;
$N_p$ is the proton density;
$T_p$ is the proton temperature;
$V_p$ is the proton bulk velocity;
$\phi_{VB}$ is the angle between the mean solar wind velocity and magnetic field;
$f_{cp}$ is the proton gyrofrequency;
$r_{cp}$ is the proton gyroradius;
$d_p$ is the proton inertial length;
$\beta$ is the ratio of proton thermal to magnetic pressure;
$\delta B / \langle B \rangle$ is the relative amplitude of magnetic fluctuations.} 

\tablenotetext{b}{Magnetic compressibility is defined as $C_\parallel=\langle\frac{|\delta B_\parallel(f_{sc})|^2}{|\delta B_\parallel(f_{sc})|^2+|\delta B_\perp(f_{sc})|^2}\rangle_f$, where $\delta{B}_\parallel$ and $\delta B_\perp$ are magnetic fluctuations parallel and perpendicular to $\hat{\mathbf{b}}_0$. $\langle\cdot\rangle_f$ denotes the frequency average over $0.001-0.05$ Hz. The spectral index $\alpha_B$ is obtained by applying a least-squares fit to the spacecraft-frame magnetic power spectrum over $0.001-0.05$ Hz, using three-point smoothing. }

\tablenotetext{c}{The tetrahedron quality factor (TQF) characterizes the four Cluster spacecraft configuration. Events 1, 2, 3, and 6 (TQF$>0.8$) were used for timing analysis. Event 5 was excluded due to insufficient magnetic-field resolution.}
\end{table*}

\bibliography{sample701}{}
\bibliographystyle{aasjournalv7}




\end{document}